\documentclass{Interspeech}

\interspeechcameraready

\title{Towards Better Disentanglement in Non-Autoregressive Zero-Shot Expressive Voice Conversion}

\author[affiliation={1}]{Seymanur}{Akti}
\author[affiliation={1}]{Tuan Nam}{Nguyen}
\author[affiliation={1,2}]{Alexander}{Waibel}

\affiliation{Interactive Systems Lab}{Karlsruhe Institute for Technology}{Germany}
\affiliation{}{Carnegie Mellon University}{USA}
\email{seymanur.akti@kit.edu, tuan.nguyen@kit.edu, alexander.waibel@kit.edu}
\keywords{speech synthesis, expressive voice conversion}

\usepackage{comment}
\usepackage{multirow}
\usepackage[table,xcdraw]{xcolor}
\usepackage{graphicx} 

\begin{document}

\maketitle

\begin{abstract}
    
    Expressive voice conversion aims to transfer both speaker identity and expressive attributes from a target speech to a given source speech. In this work, we improve over a self-supervised, non-autoregressive framework with a conditional variational autoencoder, focusing on reducing source timbre leakage and improving linguistic-acoustic disentanglement for better style transfer. To minimize style leakage, we use multilingual discrete speech units for content representation and reinforce embeddings with augmentation-based similarity loss and mix-style layer normalization. To enhance expressivity transfer, we incorporate local F0 information via cross-attention and extract style embeddings enriched with global pitch and energy features. Experiments show our model outperforms baselines in emotion and speaker similarity, demonstrating superior style adaptation and reduced source style leakage.
\end{abstract}

\section{Introduction}
Voice conversion (VC) approaches aim to transform a source audio by transferring speaker characteristics from a target audio while preserving the source content. Conventional VC models perform well in replicating speaker identity but struggle when the target speech is highly expressive. Expressive voice conversion (EVC) expands on this by capturing both speaker and expressive cues, such as emotions, intensity, and pitch, during synthesis. This allows to not only generate read speech but also replicate the emotional nuances of the target speaker. EVC can be applied in dialogue systems such as~\cite{waibe112005chil,schmidt2017towards} to improve human-robot interactions or speech translation pipelines~\cite{waibel2023face, ahmad2024findings, waibel2012simultaneous}, ensuring that the emotions and expressions from the source language are retained in the target language speech~\cite{barrault2023seamless,song2023styles2st}. 

EVC was tackled as a supervised sequence-to-sequence task~\cite{zhou2021limited}, however, due to the challenge of creating parallel emotional speech corpora, many methods explored non-parallel synthesis~\cite{gao2018nonparallel,cao2020nonparallel}. A common strategy is disentangling linguistic and acoustic information in a self-supervised manner, then recombining the source’s linguistic features with the target’s acoustic attributes~\cite{kreuk2021textless}. However, a persistent challenge is source timbre leakage, where residual speaker timbre from the source speech degrade style transfer. To address this, some studies introduce an information bottleneck in the linguistic encoder to suppress residual acoustic information~\cite{lee2023hiervst,ning2023expressive}.

In this work, we enhance the information bottleneck to improve disentanglement and reduce source speaker leakage. We adopt a non-parallel, self-supervised speech generation model based on a conditional variational autoencoder, inspired by VITS~\cite{kim2021conditional} and FreeVC~\cite{li2023freevc}. Our system uses self-supervised speech representations from mHuBERT-147~\cite{boito2024mhubert} as input, leveraging its discrete speech units to eliminate non-linguistic information more effectively than continuous representations through quantization~\cite{akti2024voice}. Additionally, its multilingual speech units enable cross-lingual EVC, making it particularly valuable for speech translation pipelines. To our knowledge, this is the first use of mHuBERT-147 in a VC setup, combining the benefits of both discrete speech units and multilinguality.
To further improve disentanglement, we introduce a perturbation-based similarity loss to minimize variation in the content embedding distribution and integrate mixed-layer normalization from~\cite{huang2022generspeech} to enhance the linguistic-acoustic disentanglement. We represent both speaker identity and emotional cues in a single global style embedding, using ECAPA-TDNN~\cite{desplanques2020ecapa} complemented by global pitch and energy information. Additionally, we add a local F0 encoder for a better prosodic similarity to the target speech. We provide speech samples on the demo page.\footnote{\url{https://seymanurakti.github.io/evc/}}. 

\begin{figure*}[h!]
\centering

\includegraphics[width=\textwidth]{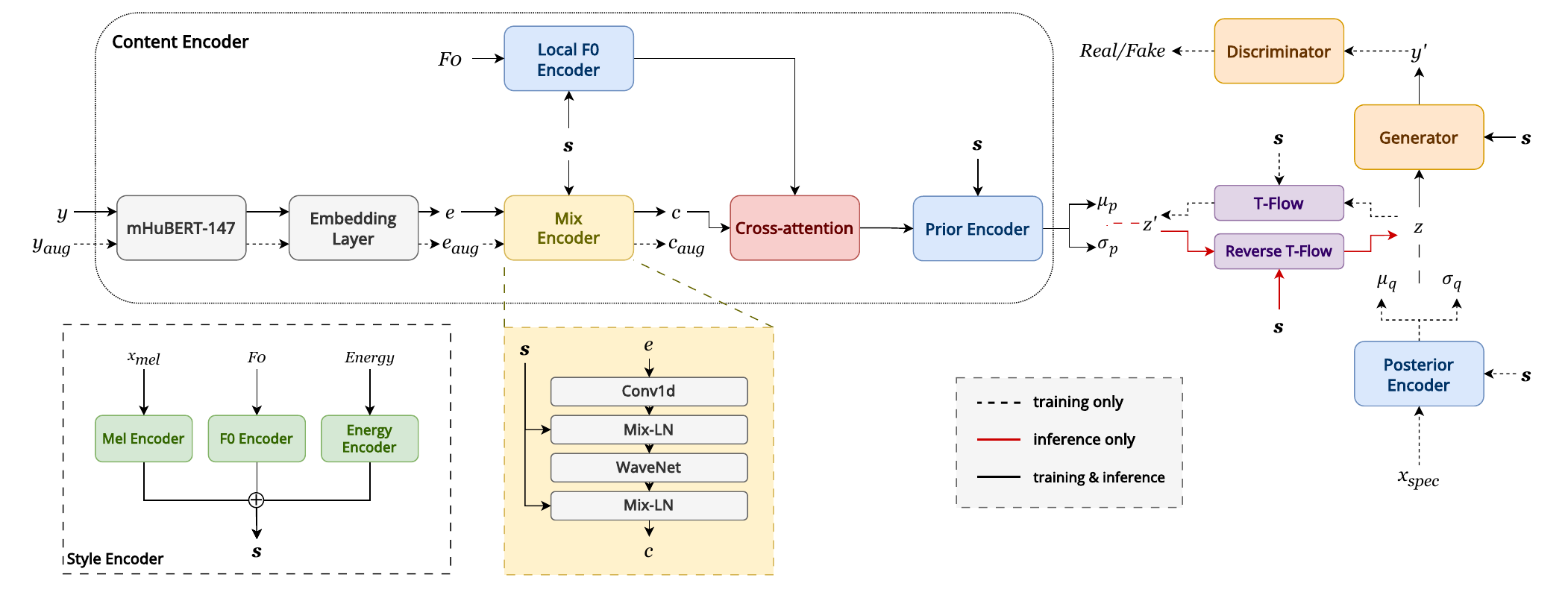}
\caption{Overall architecture of the proposed system.}
\label{fig:architecture}
\end{figure*}

\section{Related Work}

EVC is a speech-to-speech task, where early approaches often relied on auto-regressive models trained with parallel speech data~\cite{zhou2021limited, zhou2022emotion}. These models typically required text supervision. To reduce reliance on transcripts, \cite{kreuk2021textless} proposed a textless approach that extracts discrete speech units directly from audio and learns a translation network between emotion classes. Given that obtaining original parallel data is challenging, some studies explored using synthesized parallel data for accent conversion~\cite{nguyen2022accent,nguyen2025improving}. 

Meanwhile, in EVC, many recent methods have adopted self-supervised speech reconstruction techniques to remove the need for aligned data. One of the earliest studies in this area, \cite{gao2018nonparallel}, disentangles emotion-invariant and emotion-variant features and uses an autoencoder to synthesize speech from these features. \cite{rizos2020stargan} employs StarGAN for the EVC task while \cite{cao2020nonparallel} introduces variational autoencoders with non-autoregressive speech synthesis for EVC. \cite{schnell2021emocat} applies a VAE for language-agnostic EVC with limited data. \cite{baas2023voice} leverages self-supervised speech representations and a k-nearest neighbors model for feature retrieval. \cite{qu2023disentangling} uses discrete speech units as linguistic inputs and a prosody encoder for acoustic inputs, synthesizing speech with an auto-regressive decoder. The most similar works to ours are conditional VAE based methods adapted for the style conversion task where~\cite{lee2023hiervst} uses a hierarchical adaptive generator for generating the waveform, ~\cite{guo2023using} adds the style consistency loss for a better style transfer, ~\cite{guo2024xe} uses a similar architecture to ~\cite{lee2023hiervst} and adapts it for jointly trained TTS and cross-lingual EVC and~\cite{li2023zse} introduces prosody extraction and fusing methods for improving VITS for EVC task. 

 \section{Methodology}

We follow the architecture of FreeVC~\cite{li2023freevc} and adapt it for EVC with significant modifications as overall architecture illustrated in Fig.~\ref{fig:architecture}. The aim is to learn two distributions: one for linguistic features $p = N(\mu_p, \sigma_p)$ and one for spectrogram features $q = N(\mu_q, \sigma_q)$. During training, the normalizing flow maps the posterior distribution $q$ to the prior distribution $p$, while the decoder generates speech from samples of $q$. The content encoder aims to capture only linguistic information, with acoustic features injected via style embeddings through conditional layers. During inference, the reverse flow generates style-injected representations from linguistic features, allowing speech synthesis that preserves the source content while adopting the target style.

\subsection{Content Encoder}
First, mHuBERT-147~\cite{boito2024mhubert} units are extracted from waveform $y$ and quantized. Then, unit embeddings $e$ are obtained through an embedding layer and are processed through the Mix Encoder, which helps to produce style-agnostic content embeddings by applying mixed-layer normalization~\cite{huang2022generspeech} with scale and bias vectors derived from the mixed style embeddings as:

\begin{equation}
\label{eq:gamma}
    \gamma_{mix}(s) = \lambda \gamma(s) + (1-\lambda) \gamma(\widetilde{s})   
\end{equation}
\begin{equation}
\label{eq:beta}
     \beta_{mix}(s) = \lambda \beta(s) + (1-\lambda) \beta(\widetilde{s}) 
\end{equation}
\begin{equation}
\label{eq:mix-ln}
    MixLN(e,s) = \gamma_{mix}(s) \times LN(e) + \beta_{mix}(s)
\end{equation} where $\widetilde{s}$ is batch-wise shuffled style embeddings and $\lambda$ is a parameter from Beta distribution. This introduces random style information from other samples in the batch, injecting some noise, reducing the dependency of content embeddings on their original style embeddings and making them more style-agnostic. Our experiments show that Mix-LN also improves content-independence in style embeddings implicitly making them time-invariant. This is particularly useful for mitigating the train-inference mismatch, where training samples share the same content for source and target speech, whereas this alignment does not hold during inference.

Concurrently, the Local F0 Encoder extracts frame-based F0 embeddings from the audio’s F0 contours which share the same sampling rate (320) as the content embeddings $c$. F0 embeddings are then fused with the content embeddings via multi-head cross-attention, with content embeddings as query and F0 embeddings as key and value. This more effectively captures the pitch flow of the target compared to re-normalizing the source F0 and fusing via summation, as done in~\cite{huang2022generspeech, lee2023hierspeech++}. It also allows the target F0 to be directly used as the pitch input during inference—regardless of any length differences. Finally, the prior encoder generates the distribution $p(z|c)$ from the F0-enriched content embeddings.
 
Additionally, we introduce a perturbation-based similarity loss to improve the quality of content embeddings. We apply augmentation via Parselmouth\footnote{\url{https://github.com/YannickJadoul/Parselmouth}}, modifying the original audio by reducing its pitch range to create less expressive and more uniform speech, and applying pitch shifting to augment speaker identity. The similarity loss then ensures that content embeddings from the original and augmented samples remain close, reducing their dependency on non-linguistic variations. The similarity loss function is formally defined in Eq.~\ref{eq:loss_con}.
\begin{equation}
\label{eq:loss_con}
L_{sim} = (1 - cos(e, e_{aug})) + (1 - cos(c, c_{aug}))
\end{equation}

\subsection{Posterior Encoder}

Posterior Encoder takes the linear spectrogram $x_{spec}$ as the input and generates the posterior distribution $q(z|x_{spec})$, sharing the same architecture with the Prior Encoder with convolution layers and projection layer for learning distribution parameters. In order to make the posterior and prior distributions closer, The KL divergence loss is calculated as given in Eq.~\ref{eq:kl_loss}.
\begin{equation}
\label{eq:kl_loss}
    L_{kl} = KL(q(z|x_{spec})||p(z|c))
\end{equation}

\subsection{Normalizing Flow}

The normalizing flow layer learns the mapping from the posterior distribution to the prior distribution during training. We use a Transformer-based normalizing flow, following the approach in~\cite{lee2023hierspeech++}, which has demonstrated superior performance compared to convolution-only normalizing flows due to its ability to capture longer time dependencies~\cite{kong2023vits2}. During inference, the content representation of the source speech is mapped to the posterior distribution using reverse normalizing flow to be decoded by Generator.

\subsection{Waveform Synthesizer}

For speech generation, we utilized the HiFi-GAN vocoder~\cite{kong2020hifi}. The Generator generates speech waveforms through a series of upsampling layers, while the Discriminator aims to distinguish between real and generated waveforms. The synthesizer employs three loss functions similar to conventional generative adversarial networks, as defined in Eq.~\ref{eq:loss_advg}-\ref{eq:loss_fm}.

\begin{equation}
\label{eq:loss_advg}
    L_{adv}(G) = \mathbb{E}_z[(D(G(z))-1)^2]
\end{equation}
\begin{equation}
\label{eq:loss_advd}
    L_{adv}(D) = \mathbb{E}_{(y,z)}[(D(y)-1)^2 + (D(G(z)))^2]
\end{equation}
\begin{equation}
\label{eq:loss_fm}
    L_{fm}(G) = \mathbb{E}_{(y,z)}[\sum_{l=1}^T \frac{1}{N_l} \lVert D^l(y) - D^l(G(z))\rVert_1]
\end{equation}

Also, a reconstruction loss is calculated between the original and generated mel spectrograms as shown in Eq.~\ref{eq:loss_rec}.
\begin{equation}
\label{eq:loss_rec}
    L_{rec} = \lVert x_{mel} - \hat{x}_{mel} \rVert_1
\end{equation}

\subsection{Style Encoder}

We use the ECAPA-TDNN model~\cite{desplanques2020ecapa} for style encoding. Unlike style encoders relying solely on mel spectrograms, we also incorporate F0 and energy contours. We extract 512-dimensional embeddings from mel spectrograms, F0, and energy contours, then fuse them using a trainable weighted summation, as defined in Eq.~\ref{eq:style}.

\vspace{-0.5cm}

\begin{equation}
\label{eq:style}
    s = \lambda_{mel} . e_{mel} + \lambda_{f0} . e_{f0} + \lambda_{energy} . e_{energy}
\end{equation}

Style embeddings are then used to condition all encoders, injecting style information throughout the process. The overall loss function is given in Eq.~\ref{eq:loss}.
\begin{equation}
\label{eq:loss}
    L = L_{adv}(G) + L_{adv}(D) + L_{fm}(G) + L_{kl} + L_{rec} + L_{sim}
\end{equation}

\begin{table*}[h!]
\setlength{\tabcolsep}{4pt}
\centering
\caption{Comparative results. N$\rightarrow$O indicates conversion from neutral to the other emotions. (overall) is for cross-emotion conversion.}
\label{tab:results}
\begin{tabular}{@{}lcccc|cc|ccc@{}}
\toprule
\multicolumn{1}{c}{\multirow{2}{*}{Models}} & \multicolumn{4}{c|}{ESD}                                                                                      & \multicolumn{2}{c|}{Expresso}     & \multicolumn{3}{c}{LibriTTS}                       \\ \cmidrule(l){2-10} 
\multicolumn{1}{c}{}                         & WER $\downarrow$             & ECA (N $\rightarrow$ O) $\uparrow$ & ECA (O $\rightarrow$ N) $\uparrow$ & ECA (overall) $\uparrow$   & \multicolumn{1}{c}{SECS $\uparrow$ } & \multicolumn{1}{c|}{EECS $\uparrow$ } & WER $\downarrow$             & SECS $\uparrow$             & EER $\downarrow$           \\ \midrule
\multicolumn{1}{l|}{Consistency-VC}         & \textbf{4.71\%} & \textbf{78.3\%}                     & 72.6\%                              & \textbf{77.0\%} & 66.6\%          & 79.4\%          & -               & -               & -              \\
\multicolumn{1}{l|}{X-E-Speech}             & 5.69\%          & 69.6\%                              & 74.1\%                              & 68.9\%          & 67.1\%          & 78.7\%          & -               & -               & -              \\
\multicolumn{1}{l|}{Ours (ESD only)}        & 10.44\%         & 76.3\%                              & \textbf{78.8\%}                     & 76.5\%          & \textbf{67.9\%} & \textbf{81.2\%} & -               & -               & -              \\ \midrule
\multicolumn{1}{l|}{Hierspeech++}           & \textbf{5.01\%} & 35.7\%                              & 45.0\%                              & 37.0\%          & 73.1\%          & 82.0\%          & \textbf{3.48\%} & 80.3\%          & 14.6\%         \\
\multicolumn{1}{l|}{Ours}                   & 7.98\%          & \textbf{81.2\%}                     & \textbf{73.8\%}                     & \textbf{78.9\%} & \textbf{81.2\%} & \textbf{85.3\%} & 8.84\%          & \textbf{83.2\%} & \textbf{7.4\%} \\ \bottomrule

\end{tabular}
\end{table*}

\vspace{-0.1cm} 

\section{Experiments and Results}

For training, we used a combination of LibriTTS-100~\cite{zen2019libritts}, ESD~\cite{zhou2021seen} (English only), subset of GigaSpeech~\cite{chen2021gigaspeech}, and Expresso~\cite{nguyen2023expresso}. All datasets are English and total duration is around 228 hours with more than 920 speakers. We used 2 NVIDIA A600 GPUs for training with batch size of 64 for 1M steps. For the ablation study, we trained the models for 300k steps.  

\begin{table*}[t!]
\centering
\caption{Ablation results. N$\rightarrow$O indicates conversion from neutral to other emotions. (overall) indicates cross-emotion conversion.}
\label{tab:ablation_study}
\begin{tabular}{@{}lccc|ll|ccc@{}}
\toprule
\multicolumn{1}{c}{\multirow{2}{*}{Models}}   & \multicolumn{3}{c|}{ESD}                                                           & \multicolumn{2}{c|}{Expresso}                        & \multicolumn{3}{c}{LibriTTS}                       \\ \cmidrule(l){2-9} 
\multicolumn{1}{c}{}                          & WER $\downarrow$             & ECA (N $\rightarrow$ O) $\uparrow$ & ECA (overall) $\uparrow$   & \multicolumn{1}{c}{SECS $\uparrow$ } & \multicolumn{1}{c|}{EECS $\uparrow$ } & WER $\downarrow$             & SECS $\uparrow$             & EER $\downarrow$            \\ \midrule
\multicolumn{1}{l|}{w/o F0 cross-attention}   & 8.82\%          & 68.3\%                                         & 64.0\%          & 79.6\%                   & 83.6\%                    & 7.42\%          & 82.0\%          & 11.1\%         \\
\multicolumn{1}{l|}{w/o Mix-LN}               & 10.45\%         & 72.9\%                                         & 74.6\%          & 78.8\%                   & 84.5\%                    & 19.83\%         & 79.4\%          & 13.9\%         \\
\multicolumn{1}{l|}{w/o $L_{sim}$}           & 9.90\%          & \textbf{77.1\%}                                & 73.0\%          & 80.1\%                   & \textbf{85.2\%}           & 10.35\%         & 81.9\%          & 9.7\%          \\
\multicolumn{1}{l|}{w/o global F0 and energy} & 9.38\%          & 76.8\%                                         & 75.4\%          & 80.3\%                   & 84.4\%                    & 9.58\%          & 80.4\%          & 10.3\%         \\
\multicolumn{1}{l|}{w/ MMS features}          & \textbf{5.01\%} & 62.2\%                                         & 60.5\%          & 79.1\%                   & 83.4\%                    & \textbf{5.21\%} & 80.3\%          & 16.5\%         \\ \midrule
\multicolumn{1}{l|}{Proposed}                 & 9.25\%          & \textbf{77.1\%}                                & \textbf{76.5\%} & \textbf{80.6\%}          & 84.9\%                    & 9.51\%          & \textbf{82.3\%} & \textbf{9.3\%} \\ \bottomrule
\end{tabular}
\end{table*}
\subsection{Evaluation}

For evaluation, we used test sets of ESD~\cite{zhou2021seen}, Expresso~\cite{nguyen2023expresso}, and LibriTTS~\cite{zen2019libritts}. For ESD, the source and target samples are from same speaker with same content and different emotions. In other two datasets, source and target had different speakers and content. For objective evaluation, we use several metrics as follows:
\begin{itemize}
 \item \textbf{WER:} We use Whisper-Large-3~\cite{radford2023robust} with text normalization.
 \item \textbf{SECS:} Speaker embedding cosine similarity computed between synthesized and target audio using Resemblyzer\footnote{\url{https://github.com/resemble-ai/Resemblyzer}}.
 \item \textbf{EECS:} Emotion embedding cosine similarity between synthesized and target audio using Emotion2Vec+~\cite{ma2023emotion2vec}.
 \item \textbf{ECA:} Emotion classification accuracy calculated on the synthesized samples using Emotion2Vec+.
 \item \textbf{EER:} Equal error rate computed with a speaker verification model~\cite{desplanques2020ecapa}, with synthesized sample as query, source speech as negative, and target as positive. A lower EER signals better target speaker matching and less source speaker leakage.
\end{itemize}

We use three subjective metrics: naturalness MOS (nMOS) for speech quality, speaker MOS (sMOS) for speaker similarity, and emotion MOS (eMOS) for emotion accuracy, all rated on a 1–5 scale. We use ESD test set samples for ESD-only models and RAVDESS~\cite{livingstone2018ryerson} for zero-shot EVC for human evaluation. 15 users participated in evaluation and users were given 10-15 samples per model across different emotions.

\subsection{Style Transfer Results}

We compare our approach with three models with VAE architecture and a style encoder for copying expressive style information by integrating emotional datasets in the training. 
For a fair comparison, we report results in two settings based on the training data: (1) ESD-only setting, comparing against X-E-Speech~\cite{guo2024xe} (trained on ESD) and Consistency-VC~\cite{guo2023using} (trained on ESD and VCTK); and (2) multi-dataset setting, comparing against Hierspeech++~\cite{lee2023hierspeech++}, which is trained on a larger dataset.

Table~\ref{tab:results} shows our model surpasses Hierspeech++ in emotion copying across all test sets, achieving higher SECS for both seen (Expresso) and unseen (LibriTTS) speakers, while lower EER on LibriTTS suggests better source speaker identity removal. In the ESD-only comparison, we outperform X-E-Speech and Consistency-VC in converting emotional speech to neutral, proving effective in eliminating source style. It also achieves higher ECA than X-E-Speech in 'neutral to others' and 'overall' but is slightly surpassed by Consistency-VC. In the zero-shot setting with unseen speakers from Expresso, our model shows slightly better emotion and speaker transfer. However, its WER is higher in both setups, likely due to discrete speech units reducing speaker information but also eliminating some linguistic cues, leading to pronunciation artifacts~\cite{van2022comparison, akti2024voice}.
 

\begin{table}[]
\caption{Subjective evaluation results.}
\label{tab:mos_results}
\begin{tabular}{@{}lccc@{}}
\toprule
                 & nMOS                 & eMOS                 & sMOS                 \\ \midrule
GT               & 4.40 ± 0.15          & 4.17 ± 0.18          & 4.60 ± 0.12          \\
Consistency-VC   & \textbf{4.25 ± 0.15} & 3.88 ± 0.19          & \textbf{4.73 ± 0.08} \\
X-E-Speech       & 4.12 ± 0.17          & 3.88 ± 0.16          & 4.63 ± 0.10          \\
Ours (ESD only)  & 3.88 ± 0.17          & \textbf{4.00 ± 0.16} & 4.50 ± 0.13          \\ \midrule
GT               & 4.50 ± 0.12          & 4.27 ± 0.16          & 4.66 ± 0.12          \\
Hierspeech++     & 2.68 ± 0.17          & 2.94 ± 0.16          & 2.97 ± 0.19          \\
Ours (w/ F0 sum) & \textbf{3.41 ± 0.17} & 3.51 ± 0.16          & \textbf{3.74 ± 0.17} \\
Ours (w/ MMS)    & \textbf{3.41 ± 0.17} & 3.46 ± 0.17          & 3.58 ± 0.18          \\
Ours (proposed)  & 3.37 ± 0.19          & \textbf{3.72 ± 0.17} & 3.65 ± 0.18          \\ \bottomrule
\end{tabular}
\end{table}

The subjective metric results at Table~\ref{tab:mos_results}, support the objective evaluation, as we surpass others in emotion transfer capability for both setups. For speaker transfer, we perform better for zero-shot, but others perform better for ESD version. Given the nMOS scores of our model exceeding 3, it is possible to claim that the audio remains reasonably natural and intelligible, despite the higher WER compared to other methods. 

\subsection{Ablation Study}

For assessing the effect of each contribution on EVC, we conducted an ablation study, as shown in Table~\ref{tab:ablation_study}. Results show that replacing F0 injection via summation with cross-attention significantly enhances style transfer. The addition of Mix-LN improves both emotion and speaker similarity, and as reflected in higher ECA (O $\rightarrow$ N) and lower EER scores, Mix-LN effectively reduces source style leakage. Furthermore, the lower difference between WERs of ESD (source and target has the same content) and LibriTTS (source and target has different content) suggests that Mix-LN mitigates the train-inference mismatch caused by content differences between the source and target speech, possibly by reducing content leakage in the style embeddings. To investigate this, we measured the intra-speaker cosine similarity of style embeddings from the style encoders of both the proposed model and the model without Mix-LN. Ideally, these embeddings should be content-agnostic, meaning similarity should remain high for the same speaker regardless of the content. For unseen speakers in LibriTTS, the average cosine similarity increased from 0.719 to 0.748 with Mix-LN, demonstrating improved disentanglement in style embeddings.

Addition of $L_{sim}$ improves unseen speaker conversion scores on LibriTTS,  indicating reduced speaker identity leakage. The higher overall ECA score further suggests that $L_{sim}$ enhances cross-emotional conversion by effectively removing source speech style. Comparing content features from MMS~\cite{pratap2024scaling} and mHuBERT, using discrete speech units as content embeddings instead of continuous ones eliminates leaked acoustic information from the source but degrades quality, as reflected in higher WER scores. Lastly, incorporating global F0 and energy embeddings improves the style embeddings and result in better speaker and emotion transfer across all metrics.

\begin{table}[]
\caption{Cross-lingual expressive voice conversion results.}
\label{tab:cross-lingual}
\setlength{\tabcolsep}{3pt}

\resizebox{\columnwidth}{!}{%
\begin{tabular}{@{}lccc|ccc@{}}
\toprule
\multirow{2}{*}{}                   & \multicolumn{3}{c|}{English to German}              & \multicolumn{3}{c}{German to English}                \\ \cmidrule(l){2-7} 
                                    & WER             & SSIM            & ECA             & WER              & SSIM            & ECA             \\ \midrule
\multicolumn{1}{l|}{Hierspeech++}   & 4.51\%          & 72.8\%          & 61.7\%          & \textbf{7.03}\%           & 65.5\%          & 15.3\%          \\
\multicolumn{1}{l|}{Consistency-VC} & \textbf{3.27\%} & 64.1\%          & 45.3\%          & 13.54\% & 73.2\%          & 25.1\%          \\
\multicolumn{1}{l|}{X-E-Speech}     & 4.77\%          & 59.8\% & 51.9\% & 43.70\%          & 76.1\%          & 47.7\%          \\
\multicolumn{1}{l|}{Ours}           & 10.59\%         & \textbf{74.8\%} & \textbf{76.1\%} & 30.84\%          & \textbf{76.8\%} & \textbf{50.7\%} \\ \bottomrule
\end{tabular}
}

\end{table}

\subsection{Cross-Lingual Style Transfer}

Considering that the style encoder is content-agnostic, and linguistic features were extracted from a multi-lingual model, our model can perform cross-lingual VC (XVC) even when trained solely on English data. In order to evaluate it, we used ESD (English) and EmoDB~\cite{burkhardt2005database} (German) to perform neutral-to-emotional conversions across different languages. We choose German to demonstrate the performance on an unseen language for all models. We use XVC version for Consistency-VC which is trained on multilingual data including ESD. As shown in Table~\ref{tab:cross-lingual}, our model more effectively preserves emotion and speaker identity, even for German speakers unseen during training. However, the WER scores reveal a significant drop in intelligibility when source language is not included in training, observed on German to English conversions for all models.

\section{Conclusion}

In this work, we proposed a novel zero-shot EVC framework that enhances linguistic and acoustic feature disentanglement to particularly reduce source style leakage. Our approach integrates F0 injection with cross-attention, Mix-LN, mHuBERT-147 units, perturbation-based similarity loss, and style embeddings enriched with F0 and energy contours in a novel framework. Experimental results demonstrate that proposed framework improves disentanglement, mitigates source style leakage more effectively than baselines, and achieves superior emotion transfer while preserving speaker similarity. Future work will focus on enhancing intelligibility and improving cross-lingual performance through multilingual training data.

\section{Acknowledgements}
The authors gratefully acknowledge support from the German Federal Ministry
of Education and Research (BMBF) under grant 01EF1803B
(RELATER), European Union’s Horizon research and innovation programme under grant 101135798 (Meetween), and KIT Campus
Transfer GmbH (KCT) staff in accordance with the collaboration with Carnegie-AI.

\bibliographystyle{IEEEtran}
\bibliography{ref}

\end{document}